# Does Information Have Mass?

LASZLO B. KISH [1], CLAES G. GRANQVIST [2]

[1] *Department of Electrical and Computer Engineering, Texas A&M University,*
*College Station, TX 77843-3128, USA*

[2] *Department of Engineering Sciences, The Ångström Laboratory, Uppsala University,*
*SE-75121 Uppsala, Sweden*

Does information have mass? This question has been asked many times and there are many answers even on the Internet, including on *Yahoo Answers*. Usually the answer is "no". Attempts have been made to assess the physical mass of information by estimating the mass of electrons feeding the power-guzzling computers and devices making up the Internet [1], the result being around 50 gram. Other efforts to calculate the mass of information have assumed that each electron involved in signal transfer carries one bit of information [2], which makes the corresponding mass to be about $10^{-5}$ gram. The difference between the two mass estimates is partially caused by the low energy efficiency of classical computers [3] (and please note that the efficiency of quantum computers is even worse [4]).

We will address the fundamental question of minimum mass related to a bit of information from the angles of quantum physics and special relativity. Our results indicate that there are different answers depending on the physical situation, and sometimes the mass can even be negative. We tend to be skeptical about the earlier mass estimations, mentioned above, because our results indicate that the electron's mass does not play a role in any one of them.

Finally, in a seemingly related but actually different and independent context of measured inaccuracies of the gravitation constant, we refer to experiments on the weight transient of memory devices after information writing and erasure.

## I. MASS OF THE INFORMATION CARRIER DURING COMMUNICATION

**I.1 Single electrons as information carriers**

Electrons have a finite rest mass of $m_e = 9.11 \times 10^{-31}$ kg, which is their mass at zero velocity. This is an extremely small number, but it is infinitely larger than zero which is the smallest mass a photon theoretically may have.

**I.2 Single photons**

One cannot store information with photons but it is possible to communicate information with them, either through an optical fiber or via free space. In an exploration of the fundamental lower limits of the transferred mass related to information, it is therefore more proper to use photons than electrons. In an ideal case, a single photon can carry one bit of information, for example by using the two orthogonal polarization modes to represent the two bit values. According to quantum mechanics, the energy $E_p$ of a photon is given by Planck's formula which states that

$$E_p = hf_p ,  \tag{1}$$

where $h = 6.63 \times 10^{-34}$ Js is Planck's constant and $f_p$ is the frequency of the photon. The relativistic mass $m_p$ of the photon is given by Einstein's famous equation

$$E_p = m_p c^2 ,  \tag{2}$$

where $c = 3.00 \times 10^8$ m/s is the speed of light. Equations (1) and (2) then yield that

$$m_p = \frac{h}{c^2} f_p = 2.21 \times 10^{-42} f_p \ (\text{kg}) .  \tag{3}$$

So if a single optical photon with $E_p \approx 1$ eV $\approx 1.6 \times 10^{-19}$ J carries one bit of information, the mass of the carrier of the communicated bit is $m_p \approx 10^{-36}$ kg, which is about a million times less than the electron's rest mass.

Equation (3) seemingly indicates that the mass of the information carrier can approach zero if one uses photons in the low-frequency limit where $f_p \to 0$. However, this is not so even if one were able to measure single photons of low frequency, and the reason is twofold: one may neither disregard noise (*i.e.*, the thermal radiation background) nor the uncertainty principle.

**I.3 Photons with thermal radiation background**

Each photon mode has two thermodynamic degrees of freedom. In accordance with Boltzmann's energy equipartition theorem, the mean thermal energy $E_t$ in a photon mode is

$$E_t = kT ,  \tag{4}$$

where $k$ is Boltzmann's constant and $T$ is the absolute temperature of the environment. Equation (4) is valid in the classical physical limit of thermodynamics where

$$hf_p \ll kT  \tag{5}$$

and describes the thermal radiation background, which represents the thermal noise of the electromagnetic field. This noise energy is available for each mode, including the polarization modes. In order to have a sufficiently strong signal and high signal-to-noise ratio (SNR) to achieve a small bit-error probability (BEP), the energy representing the bit value must be much greater than the corresponding noise energy, which means that

$$kT = E_t \ll E_p = hf_p .  \tag{6}$$

Equation (6) obviously contradicts Eq. (5), which is an indication that single photons cannot be used with high SNR and low BEP in the classical physical limit of thermodynamics. On the other hand the assumption $f_p \to 0$, which is needed for the photon mass $m_p$ to approach zero, implies the classical limit in Eq. (5) for any fixed temperature *T*. To avoid this contradiction in the classical limit, one must stay away from using single photons.

Looking now at a number $N_p$ of photons, the requirement of having much greater signal energy than noise energy yields

$$kT \ll N_p E_p = N_p h f_p \ . \tag{7}$$

If this is translated to the mass $m_w$ of a bit-carrying classical electromagnetic wave, one gets a frequency-independent result according to

$$m_w \gg \frac{kT}{c^2} \ , \tag{8}$$

which corresponds to $m_w \approx 4.6 \times 10^{-38}$ kg at room temperature and for zero-decibel SNR.

## I.4 The uncertainty principle

Suppose that the bits must be sent with extraordinarily high speed (which is not typical in today's optical communication owing to limitations in light sources and detectors). Then the energy–time uncertainty principle poses a new limitation which reads

$$\Delta E \ \Delta t \geq \frac{\hbar}{2} \ , \tag{9}$$

where the energy uncertainty $\Delta E$ is the root-mean-square (RMS) error in the energy measurement, the "bit duration" $\Delta t$ is the width of the time window for generating the photon or photon package with the required bit value, and $\hbar = h/2\pi$. Expression (9) can be written $\Delta E \geq \hbar/(2\Delta t)$, and the mass-uncertainty of the communicated bit is

$$\Delta m_b \geq \frac{\hbar}{2c^2 \Delta t} \ . \tag{10}$$

Expression (10) becomes significant only when the mass uncertainty is larger than the mass related to the communicated bit, because zero is the low limit of the resultant mass. A bit duration of one picosecond—corresponding to a clock frequency of 1000 GHz in a serial photonic one-bit communication port—yields

$$\Delta m_b \geq 5.9 \times 10^{-40} \text{ kg} \ . \tag{11}$$

The latter result indicates that, with a thermal background corresponding to room temperature, the effect of the uncertainty principle would begin to increase the mass of the bit only at bit frequencies exceeding $\sim 10^5$ GHz for the case of a serial one-bit communication line.

Finally, one should note that it is possible to communicate without emitting any signal energy into the communication channel, namely by modulating the statistical properties of its thermal noise or quantum noise [5]. But even in this case the noise medium carrying the information has a mass.

II. MASS OF STORED INFORMATION

Mass related to information storage is different from the problem studied above, because this medium has mass even without storing any information. The question to address is this: How large is the mass-change per bit of the storage medium if one stores information bits there? The same question could have been asked about the photonics-based communication discussed before, but then the mass is independent of the bit value and the information content unless the modulation is so fast that it introduces a signal bandwidth comparable to or greater than the frequency of the photons. We will see next that the answer to the question concerning information storage can even be *negative* mass, depending on the definition of the erased state.

We note in passing that two irrelevant terms sometimes pop up on Internet discussions sites: entropy and mass of electrons that charge up a solid-state (flash) memory. A calculation of the thermodynamic entropy does not give any clue to the correct answer, however, because the increase of entropy is relevant for the energy dissipation (which is then released to the environment) rather than for the change of the energy and mass of the system. Similarly, the electron mass does not play any role in these memories because they operate with charge neutrality (*i.e.*, their total charge is always zero, implying that the total number of electrons is constant at any information content). Similarly to any capacitor or MOS transistor in a circuit, the charges at the surface of the gate capacitance are always compensated by an equal amount of opposite charges at the other electrode of this capacitor, which is the channel between drain and source. Otherwise a pen-drive containing certain information would act like an insulating body charged up to a voltage of 1000–100000 V; it would attract dust and small pieces of papers and make one's hair stand up when being in its vicinity.

To gain a physical formulation of information storage, we first consider the erased state of the memory. In the case of *normal erasure*, one only alters the address identifying the free memory and the old information is not deleted but simply readdressed so that it belongs to the free memory, where it will eventually be written over by new information. Therefore we discuss *secure erasure* and suppose that (*i*) *all bits have zero value in the (secure-) erased memory*. Furthermore we assume that the bits do not interact and ask the question: (*ii*) *what is the energy difference between the stored values* (0 *or* 1) *of a physical bit*?

Figure 1a shows a diagram of the potential energy for a single-bit memory based on a *symmetric* double-well potential. A magnetic memory cell, for example, can approximate this situation. The height $E_1$ of the potential barrier characterizes the energy that must be invested (dissipated) to change the bit value. Flips corresponding to $0 \to 1$ and $1 \to 0$ require the same lower limit of energy, $E_1$.

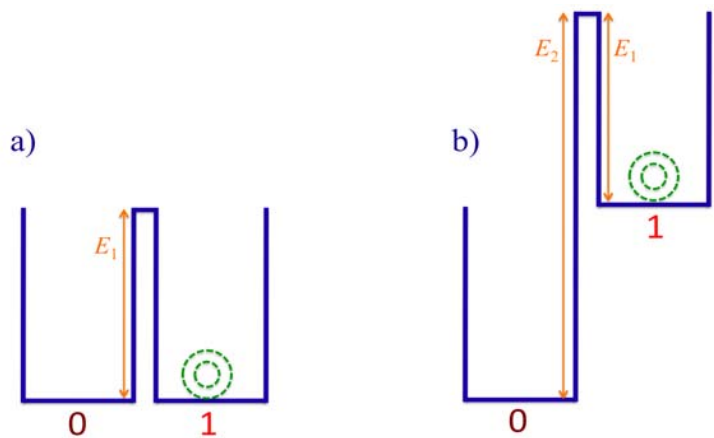

**Figure 1. Symmetric (a) and asymmetric (b) double-well-potential memories**. The two bit values correspond to identical (a) and different (b) energies, respectively.

Figure 1b illustrates a potential energy diagram of a single-bit memory based on an *asymmetric* double-well potential. A floating-gate MOSFET, which is the flash-memory element, is an example of this situation (although the energy levels for values 0 and 1 may be inverted). At one of the bit values, the floating gate—which is embedded in an insulating oxide or nitride—must be charged up by a quantum-tunneling current via the insulator in order to keep the channel of the floating-field-effect transistor in the required state. In Figure 1b, the charged position is represented by the high-energy state 1 and the discharged position by the low-energy state 0. The height $E_1$ of the potential barrier characterizes the energy that must be invested to trigger the bit-value change $1 \to 0$, and $E_2$ is needed for the $0 \to 1$ flip. These potential barriers represent thermal activation in the two directions. (Note that energies that are much larger than $E_1$ or $E_2$ are used in practice to facilitate high-speed operation via quantum-tunneling, and thus the actual energy dissipation is greater than indicated in our simplified analysis).

What is the minimum energy barrier $E_1$ needed in these devices in order for them not to "forget" their information? Supposing that nothing but thermal activation leakage takes place one gets [6,7]

$$E_1 > kT \ln\left(\frac{1}{\varepsilon_m} \frac{t_m}{\tau}\right), \tag{12}$$

where $t_m$ is the lifetime of the memory, $\varepsilon_m \ll 1$ is the probability of bit-error during this lifetime, and $\tau$ is the correlation time of the thermal fluctuations driving the thermal excitation. Assume that the memory contains *N* bits and that the goal is that the probability of even a single bit-error should be negligible during the memory's entire lifetime. Then the relation $\varepsilon_m \ll 1/N$ must hold, which leads to

$$E_1 > kT \ln\left(N \frac{t_m}{\tau}\right). \tag{13}$$

For a one-terabyte solid-state drive with ten years lifetime and $\tau = 10^{-12}$ s, one obtains $E_1 \approx 3.1 \times 10^{-19}$ J at room temperature, which corresponds to ~75 $kT$. Though this energy is dissipated by the writing operation during changing the state, the energy and the mass of the system after the operation remain the same as their original values.

On the other hand, the energy change of the asymmetric system in Figure 1b is characterized by $\Delta E = E_2 - E_1$, which is the energy needed to keep the MOSFET channel closed (or open, depending on the type of MOSFET). The BEP of that MOSFET channel is controlled by an expression that is analogous to the one for $E_1$ in expression (13) [3], and it follows that the energy change $\Delta E$ of the system must satisfy

$$\Delta E > kT \ln\left(N \frac{t_m}{\tau}\right). \tag{14}$$

In conclusion, the minimum mass change a one-bit flash memory, when flipped into bit value 1, is

$$m_f \approx \frac{kT}{c^2} \ln\left(N \frac{t_m}{\tau}\right), \tag{15}$$

where Einstein's relation in Eq. (2) has been used. This yields $m_f \approx 3.4 \times 10^{-36}$ kg at room temperature. Coincidentally, the absolute value of the mass in Eq. (15) is very similar to that of a communicated quantum bit at the chosen conditions. However, the mass in Eq. (15) is *negative* if one defines the erased memory element to be a MOSFET with charged gate!

EXPERIMENTS: NEGATIVE WEIGHT TRANSIENTS OF MEMORIES

Experiments devised to compare the weight of information storage media before and after recording/erasure have been carried out to $10^{-8}$ kg accuracy by use of a precision balance, and significant differences—the order of $10^{-6}$ kg —have been found [8,9], as discussed below.

The original goal of these experiments was not to compare weights, because the considerations in the section above make it evident that the expected differences are too small to be measurable. Instead the weight transients were discovered during tests [8] inspired by explorations of a hypothetical "Fifth Force" that conceivably complements electrostatic, strong, weak and gravitational interactions [10]. The reason behind the quest for a "Fifth Force" is the extraordinary inaccuracy of the Newtonian gravitational constant $G$, and that specific materials and environments seem to influence measurements that theoretically should offer excellent accuracy. The "Fifth-Force"-hypothesis was forwarded by Fischbach *et al.* in the 1980s but was found untenable. Nevertheless it stimulated several laboratories to carry out measurements during the 1990s with emphasis on enhancing the accuracy of *G* [11]. These efforts backfired, however, and the new and supposedly improved measurements yielded even greater inaccuracies than before for the short-range value of *G* [11]. In 1999, the Committee on Data for Science and Technology (CODATA), under the International Council for Science, decided [11] to increase the uncertainty of *G* from 128 to 1500 ppm (0.15 %). The original ~0.01 % accuracy was reclaimed the next year, but even this uncertainty is orders of magnitude beyond that of the acceptable level for other physical constants.

One of the present authors (LBK) hypothesized that the "Fifth Force" might be due to some short-range interaction between structures with similar information patterns, shared among the test bodies, and their environment [8,9]. Such an interaction goes beyond today's physics. Tests were nevertheless executed but no measurable interaction force, at a resolution of $10^{-7}$ N, was found between two identical 4-Gigabyte flash drives filled with identical noise sequences [8,9].

However, a serendipitous result emerged from these efforts to explore the "Fifth Force", namely negative weight transients [8,9] of the order of $10^{-5}$ N (corresponding to a weight of about 1 mg) during the recording or secure-erasing information (noises and periodic signals) in the Gigabyte range on flash, CD and DVD media. The relaxation of the weight transient was nearly exponential, with time constants ranging from a few minutes to 15 minutes [8,9].

An elevated temperature of the storage medium after recording/erasure could cause a lifting force due to heat convection flow and the ensuing Bernoulli force [8], but the observations were inconsistent with thermal relaxation time constants, which are much shorter than those observed. For example, the weight transient of the DVD disks typically have an order-of-magnitude longer time constant than expected for thermal relaxation. Another conceivable explanation for the weight changes is the loss of stored moisture in hygroscopic parts of the media during recording, and a corresponding mass relaxation when moisture is regained. However, the observed dependence of the time constant on the humidity of ambient air does not support this assumption. Neither are other simple explanations satisfactory, such as those involving a changed Archimedes force due to thermal expansion (which is a negligible effect).

We emphasize that the experiments were done carefully to avoid artifacts of shocks to the precision balance during placing/removing memories as well as direct contact of the balance plate with the warm flash drives. Tests against artifacts included repeated placing of the same flash drive without information recording to ascertain that the values were reproducible and time-independent in the absence of information being written into the media.

In conclusion, weight transients observed after writing or secure-erasing information into a memory have not been satisfactorily explained. Experiments with better control of the ambient conditions, for example by use of vacuum, would be desirable. Finally, we note that the observed effects offer a security application, and weight transients conceivably could serve as a "smoking gun" to evidence that a storage medium has been in recent use.